\documentstyle[12pt]{article}

\newcommand{\be}{\begin{equation}}
\newcommand{\ee}{\end{equation}}
\newcommand{\ba}{\begin{eqnarray}}
\newcommand{\ea}{\end{eqnarray}}

\begin{document}

\begin{flushright}
QMW-PH-96-28\\IC/96/276\\hep-th/9612182\\
\end{flushright}

\begin{center}

{\Large{\bf Self-Duality in $D\leq8$-dimensional \\Euclidean Gravity}}\\

B.S.Acharya\footnote{e-mail:r.acharya@qmw.ac.uk, Work supported by PPARC}\\
{\it Queen Mary and Westfield College,\\
 Mile End Rd., London. E1 4NS, UK.}\\
and\\
M. O'Loughlin\footnote{e-mail:mjol@ictp.trieste.it}\\
{\it ICTP, PO Box 586, Trieste, 34014, Italy.}
\end{center}

\begin{abstract}
In the context of $D$-dimensional Euclidean gravity,
we define the natural generalisation to $D$-dimensions of the
self-dual Yang-Mills equations, as duality conditions on the
curvature $2$-form of a Riemannian manifold. Solutions to these
self-duality equations are provided by manifolds of $SU(2), SU(3), 
G_2$ and $Spin(7)$ holonomy. The equations in eight dimensions are
a master set for those in lower dimensions. By considering gauge
fields propagating on these self-dual manifolds and embedding the
spin connection in the gauge connection, solutions to the
$D$-dimensional equations for self-dual Yang-Mills fields are
found. We show that the Yang-Mills action on such manifolds
is topologically bounded from below, with the bound saturated precisely
when the Yang-Mills field is self-dual. These results have a natural
interpretation in supersymmetric string theory.

\end{abstract}

\newpage

\section{Introduction.}

Self-duality in four dimensional Yang-Mills theory has had a remarkable
impact in physics and mathematics. It is natural to ask
if the four-dimensional equations for self-duality have an analogue in
higher dimensions. This question was considered in \cite{corr} and
a natural set of equations for self-duality in dimensions four through
eight were found (the eight dimensional equations were independently
discovered in \cite{gurs}). These equations are:
\be
F_{\mu\nu} = {1\over2}{\phi_{\mu\nu\lambda\rho}}F_{\lambda\rho}
\ee

Here, the duality operator ${\phi}_{\mu\nu\lambda\rho}$ is related
to the structure constants of the octonions \cite{corr,gurs}.
The self-dual Yang-Mills equations were considered in Euclidean
gravity by replacing the Yang-Mills curvature with the curvature
of the Riemannian $4$-manifold. These equations imply Ricci flatness and
thus their solutions obey Einstein's equations with zero cosmological constant.
It is thus natural to consider the higher dimensional equations for self-
duality \cite{corr,gurs} in the context of Euclidean gravity.
That is the purpose of this note\footnote{Equations similar to $(1)$, but
with a nonzero constant different from ${1\over 2}$ were also considered
in \cite{corr} in their investigations of generalisations of self-duality
to higher dimensions. However, these equations do not 
have such a simple interpretation in higher dimensional Euclidean gravity
or Riemannian geometry and for this reason we do not consider them
here. These equations have recently been considered in \cite{bilge}.}.

In the next section we will consider the above self-duality
equations in Euclidean gravity, by replacing the curvature of the
gauge connection with the components of the curvature $2$-form of a
$D$-dimensional Riemannian manifold. This is in analogy with the ansatz for
self-dual gravity in four dimensions \cite{eh}. In the four dimensional
case (both in Yang-Mills and Euclidean gravity) the equations for
self-duality turn out to be first order equations. We show that this
is also true for the equations considered herein. We will find that the
$D$-dimensional Riemannian manifold must have holonomy $SU(2), SU(3),
G_2$ and $Spin(7)$ in $D$=$4,6,7$ and $8$ respectively. In fact the equations
in $4,6$ and $7$ dimensions are derivable from those in $D=8$ via
dimensional reduction. 
Following this we consider gauge fields propagating on the self-dual
manifolds and by embedding the spin connection in the gauge connection,
we show that solutions to the self-duality equations in Yang-Mills
theory (1) can always be found. Once again, in analogy with the four
dimensional case, the existence of these solutions is directly related
to a topological bound on the Yang-Mills action.
Finally we discuss a natural interpretation of these results in the context of
supersymmetry and superstring theory.  
\section{Self-Duality in $D$-dimensional \newline Euclidean Gravity.}

The purpose of this section is to study the $D$-dimensional
equations for self-duality in the context of Riemmanian geometry.
We will see that self-duality has some remarkable consequences:
namely that Einstein's equations are obeyed with zero cosmological
constant.

Let $R_{ab}$ be the components of the 
curvature two-form $R$ for the $D$-dimensional oriented
Riemannian manifold ${\it M_D}$. $R$ is constructed from the
connection one-form coefficients ${\omega}_{ab}$ through:
\be
R_{ab}= d{\omega}_{ab} + {\omega}_{ac}{\wedge}{\omega}_{cb}
\ee

We are interested in studying the self-duality equations $(1)$ with
the gauge field strength replaced with the curvature $R_{ab}$.
These equations are simply:

\be
R_{ab} = {1\over2}{\phi}_{abcd}R_{cd} \equiv {^\phi R_{ab}}
\ee

We will take these equations (with the duality operator ${\phi}_{abcd}$ to
be specified shortly) to be the defining conditions for self-duality
in $D$-dimensional Euclidean gravity, with $4\leq D \leq8$.\footnote{In 
fact, it is straightforward to show that this second order
equation on the metric is actually equivalent to a first order equation.
Specifically the duality condition is equivalent to $\omega={^{\phi}\omega}$.}

We begin in $D=8$, because from here, the equations in $D=7,6,5,4$
can be derived by dimensional reduction.
In $D=8$, the natural choice, \cite{corr},
for the duality operator, ${\phi}_{abcd}$,
is the unique $Spin(7)$ invariant
four-index antisymmetric tensor which is Hodge self-dual. In fact, this tensor
can be identified with the coefficients of the self-dual $4$-form (denoted
in what follows by ${\phi}$)
which encodes the $Spin(7)$ structure of an $8$-manifold, ${\it M_8}$, with
holonomy $Spin(7)$. 

With this choice, the equations that the curvature be self-dual in the sense
of $(3)$ are a set of $7$ non-trivial constraints among the $28$ components
$R_{ab}$. Thus, if the curvature is self-dual, 
the holonomy group of ${\it M_8}$
is $21$ dimensional. From Bergers list \cite{berg}, we see that
self-duality implies that ${\it M_8}$ has holonomy $Spin(7)$. 
In fact, the equations
for self-duality are equivalent to those discussed by Bonan \cite{Bon}, who
pointed out that ${\it M_8}$ is Ricci flat, and admits a unique, 
nowhere vanishing
$4$-form, whose coefficents are precisely ${\phi}_{abcd}$. 

\subsection{$4\leq D<8$}

It was pointed out in \cite{corr} that by deleting one index, say the
eighth, in the $D=8$ case, one finds equations for $D=7$ self-duality.
Repeating the process, one gets equations in $D=6,5,4$. The $D=4$ equations
are precisely the instanton equations.
In this sense, the eight-dimensional equations are a master set in which
the lower dimensional ones are imbedded.

In the context of Euclidean gravity that we are discussing here, being
able to delete $n$-indices in the eight dimensional equations requires that
the eight-manifold is the product of an $(8-n)$-manifold,
${\it M_{8-n}}$, with
an $n$-torus or ${\it R^n}$.
The equations then reduce to a set of self-duality equations
on the $(8-n)$-manifold, which will have non-trivial holonomy. The fact that
the original eight-manifold is Ricci flat automatically implies that
${\it M_{8-n}}$ is Ricci flat also.

For $n=1$ we find that self-duality
is a set of $7$ conditions for self duality among the $21$
components of the curvature $2$-form of the $7$-manifold, ${\it M_7}$.
These equations imply that ${\it M_7}$ has holonomy $G_2$ and were first
discovered by Bonan \cite{Bon}.
In fact in this case, ${\phi_{abcd}}$ is identified
with the components of the coassociative $4$-form, which together with
its $3$-form Hodge dual encodes the $G_2$ structure of the manifold.

For $n=2$ we find a set of $7$ self-duality constraints among the $15$
components of the curvature of ${\it M_6}$. These equations imply that
${\it M_6}$ has holonomy $SU(3)$. The duality operator in this case
is identified with the components of the Hodge dual of the Kahler form.

The case $n=3$ is essentially trivial as was also noted in the gauge theory
case in \cite{corr}. This is because there exists a direction, say the
fifth coordinate, for which all components of the curvature with
indices in this direction vanish under self-duality. Thus, ${\it M_5}$
is the product of a four-manifold with a circle or the real line.
In fact, the four manifold
has holonomy $SU(2)$.

The case $n=4$ leads to the usual equations for self-dual Euclidean
gravity in four dimensions.

Explicit (non-compact) examples of metrics with the holonomy groups above are
known \cite{eh,Cal,Bry,Gibb}.
All of these may now be interpreted as gravitational instantons
which satisfy equation $(3)$.

To summarise the results of this section: self-duality in $D$-dimensional
Euclidean gravity as defined by equation $(3)$, leads to
the conclusion that the space-time manifold ${\it M_D}$
has holonomy $SU(2)$, $SU(2)\times\{1\}$, $SU(3)$,
$G_2$ and $Spin(7)$ for $D=4,5,6,7,8$ respectively. Further the duality
operator ${\phi}_{abcd}$ is identified with the components of some
fundamental, nowhere vanishing $4$-form $\phi$ on the manifold\footnote{
The $4$-form in eight dimensions can be written as
$\bar{\eta}{\gamma}^{\mu\nu\rho\sigma}{\eta}$,
where $\eta$ is a covariantly constant unit spinor.}.
All of these manifolds obey Einstein's equations with zero cosmological
constant.

\section{Self-dual Yang-Mills fields}

Using the technique of embedding the spin connection\cite{cd} in the 
gauge connection we are able to construct a self-dual 
gauge field directly from the self-dual metric of
the previous section.

Let $G_{ab}$ be the generators of one of $SU(2), SU(3), G_2$ or $Spin(7)$.
The ansatz for the gauge field (embedding the 
spin connection in the gauge group) is $A = \gamma \, G_{ab} {\omega}_{ab}$. 
The form 
index of A comes from the form index of $\omega$ while the Lie
algebra structure of A comes from that of $G$.
We easily see that 
\ba
F & = & dA + A \wedge A \\
  & = & \gamma \, G_{ab} d\omega_{ab} + \kappa \, \gamma^2 \, G_{ab} 
\omega_{ac} \wedge \omega_{cb}
\ea
where $\kappa$ is a constant that depends upon the group generated by $G_{ab}$.
In each case it is trivial to 
solve for $\gamma$ giving $F = \gamma\, G_{ab}R_{ab}$.
Duality of $F$ follows from that of $R$ and the symmetry of 
$R_{\mu\nu\lambda\rho}$ between the first pair and second pair
of indices.

As noted in the introduction, self-duality of the gauge field
in four-dimensions has a deeper meaning. Self-dual gauge fields
minimize the Yang-Mills action and the action at the minimum is a 
topological invariant. The field configuration is that 
of an instanton. It is a remarkable result and further
confirmation of the naturalness of our construction
that an entirely analogous topological bound appears in 
eight dimensions for these instantons.
The toplogical bounds for Yang-Mills field configurations in lower
dimensions are unified in the discussion of the eight dimensional
bound.

From the definition of Hodge duality and ${\phi}$-duality
in eight dimensions we can easily derive the following.

\ba
S_{YM} & = & {1\over 8.6!} {\int}_{M_8} Tr(F \wedge {^*F}) \\
       & = & {1\over 16}{\int}_{M_8} Tr(F - {^\phi F})^2 -
{1\over 8.4!}{\int}_{M_8} Tr(F \wedge F) \wedge \phi \\
       & \geq & - {1\over 8.4!}{\int}_{M_8} Tr(F \wedge F) \wedge \phi
\ea

We have used the identity
$\phi_{abij}\phi_{abkl} = 6(\delta_{ik}\delta_{jl} - \delta_{il}
\delta_{kj}) - 4\phi_{ijkl}$ to prove the equality in $(7)$.
The final line is proportional to 
the evaluated first Pontrjagin class of the
gauge bundle for which $F$ is the curvature. 
Furthermore, when $F = {^\phi F}$
the inequality is saturated. The key point is that the $\phi$-self-dual
action is a topological invariant.

In line with our previous observations on the reduction of this
eight-dimensional duality to lower-dimensional dualities we have 
corresponding reductions of this topological bound. The process of
reduction leads us finally to the well-known formula in four dimensions.
In general we have,

\be
S_{YM}^D = {1\over 4}\int_{M_D} Tr\, F^2 \geq 
-\int_{M_D} Tr(F\wedge F) \wedge \chi_{D-4}
\ee
\noindent where $\chi_{D-4}$ is proportional to the Hodge dual of
the appropriate $4$-form $\phi$ on ${\it M_D}$. That is, $\chi_{D-4}$ 
is proportional to: 1 in four dimensions; a 
trivial one-form in five dimensions; the Kahler
form in six dimensions; the Hodge dual of the 
coassociative 4-form in seven dimensions; and the 4-form itself
in eight dimensions. In each case the Yang-Mills action is bounded by a 
topological quantity and the bound is saturated when $F$ satisfies
the relevant duality equation in $D$-dimensions.

\section{Discussions}

We have shown that in eight dimensions there exists a set of master
duality equations for self-dual Yang-Mills and self-dual Euclidean
gravity in the sense defined by equations $(1)$ and $(3)$. This master equation
incorporates self-duality in lower dimensions four, five, six and seven.

Our results have an explicit link with supersymmetry. The list of
self-dual manifolds that we have presented are precisely those which
admit covariantly constant spinors. These are the the only manifolds
which provide supersymmetric vacuum solutions of theories with
local supersymmetry. This is an intriuging and unexpected surprise.

As well as being important for Einstein and Yang-Mills
theories and their supersymmetric counterparts
in eight dimensions and lower, our construction turns out to
be very natural in string theory. In the low energy effective
field theory of the heterotic string there is an anomaly cancellation 
condition,
\be
dH = tr R\wedge R - {1\over 30}Tr F\wedge F
\ee
\noindent
which is automatically satisfied by our ansatz. 
Our solutions are in fact solutions to all the 
equations of low energy heterotic string theory in 
$9+1$ dimensions when 
one takes the dilaton to be constant and the torsion to be 
zero. Hence the noncompact examples of such metrics are solitonic 
solutions to heterotic string theory\footnote{Note that by embedding the spin
connection in the gauge connection, each metric on ${\it M_D}$ automatically
gives a solution of $(1)$ in flat space.}.
More general classes of string soliton may easily be
constructed from this starting point \cite{ao}
if we now relax the restrictions
on the dilaton $\phi$ and torsion $H$ and only partially embed the 
spin connection in the gauge connection\footnote{Simple examples of
solitonic solutions with non-zero torsion are given in \cite{hsgn}.}.
New classes
of string solitons arise in these cases and  one finds
a variety of these solitons entirely analogous 
to the varieties of heterotic 5-brane solitons as 
discussed in \cite{chs}.

This new construction in field theory and string theory is intimately
linked with the existence of octonions or Cayley numbers. 
$SO(8)$ may be decomposed into $G_2 \times S^7_L \times S^7_R$
where the left and right seven spheres are left and right
multiplication by octonions\cite{gurs}. 
The eight dimensional duality condition 
is related to this decomposition in precisely the way that 
the four-dimensional duality is related to the decomposition
of $SO(4)$ into $S^3_L \times S^3_R$ where the left and right $S^3$'s
are multiplication by quaternions.

Could it be that octonions have a 
more fundamental role to play in the current attempts
to reformulate string theory? There are now several
pieces of interesting work pointing in this direction \cite{N}.

\vspace{1.5cm}
{\centerline{\bf Acknowledgement}}
B.A. would like to thank the ICTP for inviting him for the visit 
during which this work commenced and M. O'L. would like
to thank QMW college for a reciprocal invitation during which
this paper was completed.

\end{document}